\documentclass[onecollarge,natbib]{svjour2}
\bibpunct{[}{]}{,}{n}{}{,} % to get "[numbered]" references from natbib
\smartqed  % flush right qed marks, e.g. at end of proof
\usepackage{graphicx}
\journalname{Few-Body Systems (APFB2011)}
\begin{document}

\title{\boldmath
Charged bottomonium-like structures $Z_b(10610)$ and $Z_b(10650)$%\thanks{Grants or other notes
}


\author{Xiang Liu \and Dian-Yong Chen%etc.
}


\institute{Xiang Liu \at
              School of Physical Science and Technology, Lanzhou University, Lanzhou 730000,  China\\
              %              Tel.: +123-45-678910\\
%              Fax: +123-45-678910\\
              \email{xiangliu@lzu.edu.cn}           %  \\
           \and Dian-Yong Chen
           \at
              Nuclear Theory Group, Institute of Modern Physics of CAS, Lanzhou 730000, China
}

\date{Received: date / Accepted: date}

\maketitle

\begin{abstract}
The observation of two charged bottomonium-like structures $Z_b(10610)$ and $Z_b(10650)$ has stimulated
extensive studies of the properties of $Z_b(10610)$ and $Z_b(10650)$. In this talk, we briefly introduce the research status of $Z_b(10610)$ and $Z_b(10650)$ combined with our theoretical progress.

\end{abstract}

\section{Introduction}
\label{intro}

In the past 8 years, experiments have made big progress on the observations of charmonium-like or bottomonium-like state $X$, $Y$, $Z$, which have stimulated theorists' interest in revealing their underlying structures. Thus,
studying charmonium-like and bottomonium-like states is an active and important research field in hadron physics at present, which can further deepen our understanding of the properties of $X$, $Y$, $Z$ and improve our knowledge about non-perturbative QCD.

Very recently, the Belle Collaboration \cite{Collaboration:2011gj} reported the first observation of two charged bottomonium-like structure, which makes the family of bottomonium-like abundant. By analyzing the $\Upsilon(nS)\pi^\pm$ ($n=1,2,3$) and $h_b(mP)\pi^\pm$ ($m=1,2$) invariant mass spectra of $\Upsilon(5S)\to \Upsilon(nS)\pi^+\pi^-,\,h_b(mP)\pi^+\pi^-$ hidden-bottom dipion decays, Belle observed that there exist two enhancement structures around 10610 MeV and 10650 MeV, which are named as $Z_b(10610)$ and $Z_b(10650)$, where $Z_b(10610)$ and $Z_b(10650)$ are close to the thresholds of $B\bar{B}^*$ and $B^*\bar{B}^*$ respectively.

Due to the peculiarities of $Z_b(10610)$ and $Z_b(10650)$, theorists have paid more attentions to the observed novel phenomena by different approaches \cite{Bondar:2011ev,Chen:2011zv,Zhang:2011jja,Bugg:2011jr,Voloshin:2011qa,Yang:2011rp,Nieves:2011vw,
Guo:2011gu,Sun:2011uh,Chen:2011pv,Cleven:2011gp,Cui:2011fj,Navarra:2011xa}. In the following, we will introduce the theoretical progress on the study of $Z_b(10610)$ and $Z_b(10650)$.

\section{The puzzles in the hidden-bottom dipion decays of $\Upsilon(5S)$ and $Z_b$ structures}
\label{sec:1}

Before the observation of two $Z_b$ structures \cite{Collaboration:2011gj}, the Belle Collaboration measured the $e^+e^-$$\to$$\Upsilon(1S)\pi^+\pi^-$, $\Upsilon(2S)\pi^+\pi^-$ processes near the peak of the $\Upsilon(5S)$ resonance at $\sqrt{s}=10.87$ GeV \cite{Abe:2007tk}, which indicates that there exist the anomalous $\Upsilon(1S)\pi^+\pi^-$ and $\Upsilon(2S)\pi^+\pi^-$ productions, i.e, the branching ratios of
$\Upsilon(5S)\to \Upsilon(1S)\pi^+\pi^-$ and $\Upsilon(5S)\to \Upsilon(2S)\pi^+\pi^-$ are larger than the dipion-transition rates between the lower members of the $\Upsilon$ family by two orders of magnitude \cite{Abe:2007tk}.

For solving this puzzling phenomena, rescattering mechanism was proposed in Ref. \cite{Meng:2007tk}. Since $\Upsilon(5S)$ is above the threshold of $B$ meson pair, the coupled channel effect becomes important, which makes the intermediated hadronic loops constructed by $B^{(*)}/\bar{B}^{(*)}$ mesons play crucial role to understanding the anomalous $\Upsilon(1S)\pi^+\pi^-$ and $\Upsilon(2S)\pi^+\pi^-$ production \cite{Meng:2007tk}. As an alternative explanation, tetraquark state $Y_b(10890)=[bq][\bar{b}\bar{q}]$ was introduced in Ref. \cite{Ali:2009pi}. Later, authors of Ref. \cite{Ali:2009es} studied the Belle data by analyzing the dipion invariant mass spectrum and the $\cos\theta$ distribution of $Y_b$ decays into $\Upsilon(1S)\pi^+\pi^-$ and $\Upsilon(2S)\pi^+\pi^-$, and claimed that the tetraquark interpretation can well describe the anomalous rates observed by Belle \cite{Abe:2007tk}.

\begin{figure*}
\centering
  \includegraphics[width=0.20\textwidth]{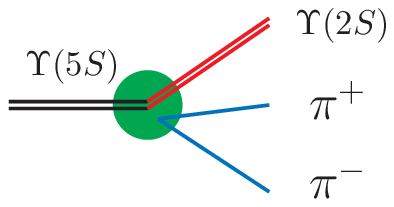}\includegraphics[width=0.2\textwidth]{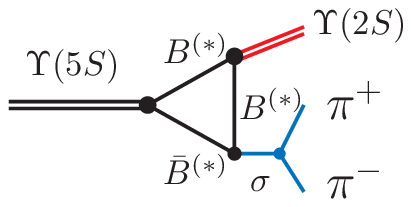}
  \includegraphics[width=0.23\textwidth]{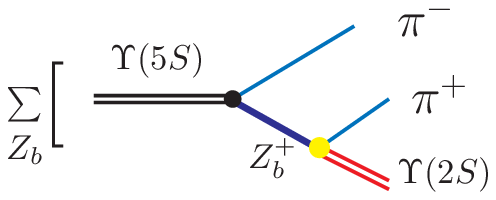}\includegraphics[width=0.23\textwidth]{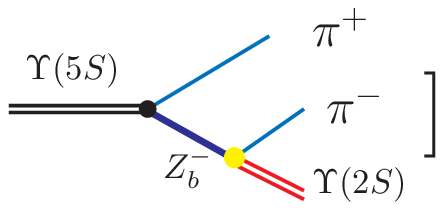}
\caption{(Color online.) The schematic diagrams of $\Upsilon(5S)\to\Upsilon(2S)\pi^+\pi^-$. The first diagram describes the direct production of $\Upsilon(2S)\pi^+\pi^-$ without the intermediated meson contribution. The second diagram is due to the rescattering mechanism \cite{Meng:2007tk}, where dipion is from intermediate $\sigma(600)$. The third and the fourth diagrams reflect two newly observed $Z_b$ structures contributing to the $\Upsilon(5S)\to\Upsilon(2S)\pi^+\pi^-$ decay.}
\label{fig:1}       % Give a unique label
\end{figure*}

Later, in Ref. \cite{Chen:2011qx} we indicated that the $\cos\theta$ distribution of $Y_b\to \Upsilon(2S)\pi^+\pi^-$ given by Ref. \cite{Ali:2009es} is not consistent with the Belle data, and proposed an alternative approach to try to explain Belle observation of $\Upsilon(5S)$ decays into $\Upsilon(1S)\pi^+\pi^-$ and $\Upsilon(2S)\pi^+\pi^-$, where the interference between the direct dipion transition and the final state interaction corresponding to the first and the second diagrams of Fig. \ref{fig:1}, respectively. Under this scenario, we can explain the anomalous rates of $\Upsilon(1S)\pi^+\pi^-$ and $\Upsilon(2S)\pi^+\pi^-$ production in $\Upsilon(5S)$ decays, especially the inverse rates $\Gamma(\Upsilon(5S)\to \Upsilon(2S)\pi^+\pi^-)>\Gamma(\Upsilon(5S)\to \Upsilon(1S)\pi^+\pi^-)$. Besides, the dipion invariant spectrum and the $\cos\theta$ distribution of $\Upsilon(5S)\to\Upsilon(1S)\pi^+\pi^-$ can be reproduced. However, the $\cos\theta$ distribution of $\Upsilon(5S)\to\Upsilon(2S)\pi^+\pi^-$ cannot be described by the scenario in Ref. \cite{Chen:2011qx}.

This fact mentioned above shows that a new puzzle appears in the $\Upsilon(5S)\to\Upsilon(2S)\pi^+\pi^-$ decay, which cannot be understood by the tetraquark state picture \cite{Ali:2009es} or rescattering mechanism \cite{Chen:2011qx}. Thus, we need to consider new mechanism involved in the $\Upsilon(5S)\to\Upsilon(2S)\pi^+\pi^-$ decay.

Since two $Z_b$ structures are from the hidden-bottom dipion decays of $\Upsilon(5S)$ \cite{Abe:2007tk},
we realized that there exists the extra intermediate $Z_b(10610)$ and $Z_b(10650)$ contributions to the $\Upsilon(5S)$ decays just depicted by the last two diagrams in Fig. \ref{fig:1}. Thus, in Ref. \cite{Chen:2011zv} we included all mechanisms listed in Fig. \ref{fig:1} to redo the analysis of the $\Upsilon(5S)\to \Upsilon(2S)\pi^+\pi^-$ decay.

\begin{figure*}[htb]
\centering
  \includegraphics[width=0.46\textwidth]{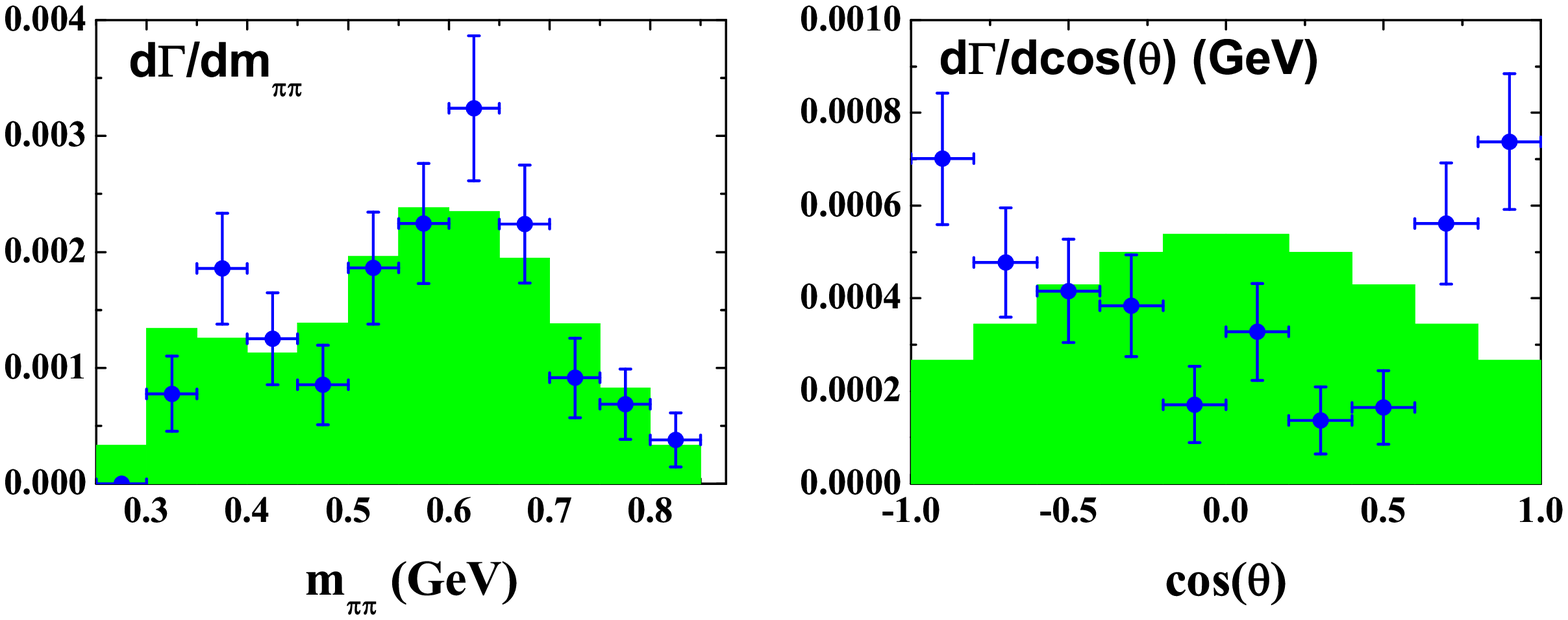}\includegraphics[width=0.46\textwidth]{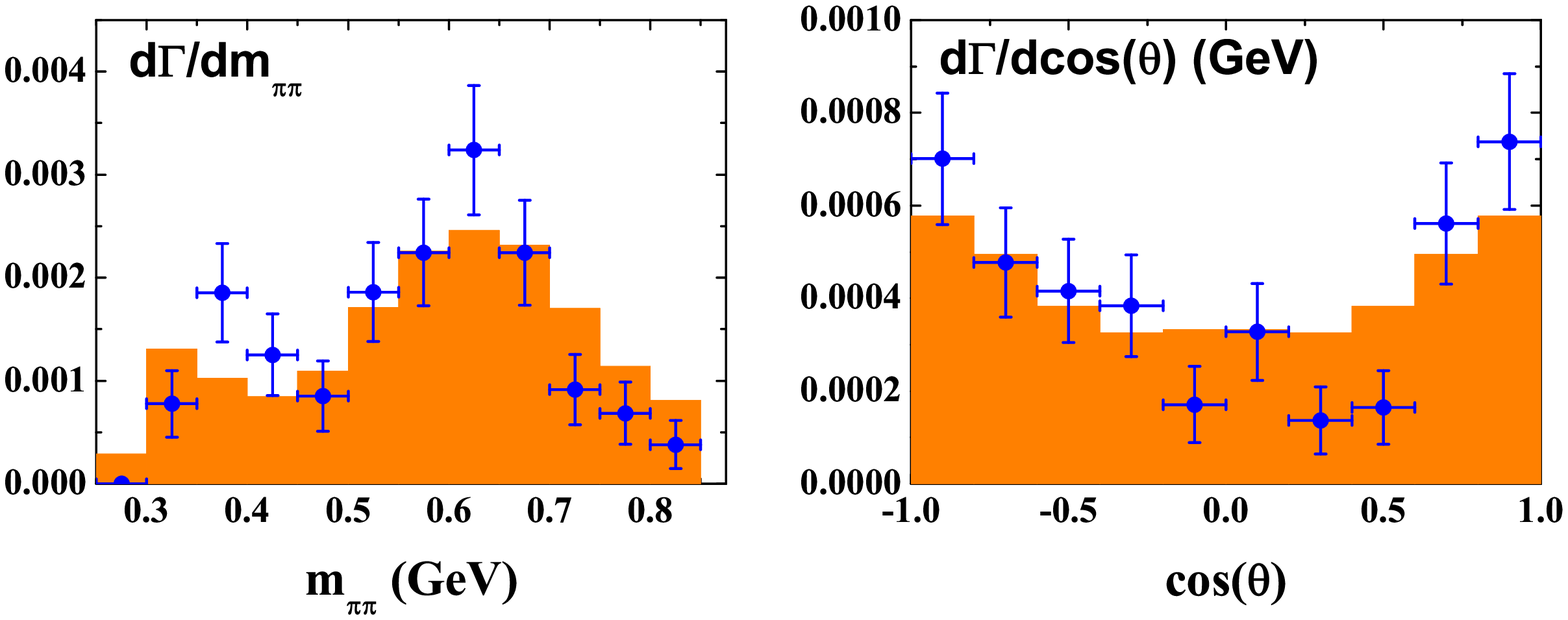}
\caption{(Color online.) The dipion invariant mass spectrum and the $\cos\theta$ distribution of the $\Upsilon(5S)\to \Upsilon(2S)\pi^+\pi^-$ decay. Here, the dots with error bars are from Belle measurement, while the histograms
are the theoretical results \cite{Chen:2011zv}. The first two diagrams or the last two diagrams are the results without or with the intermediate $Z_b$ contribution to the $\Upsilon(2S)\pi^+\pi^-$ decay. }
\label{fig:2}       % Give a unique label
\end{figure*}

The results presented in Fig. \ref{fig:2} indicate that including the intermediate $Z_b(10610)$ and $Z_b(10650)$ contribution can produce the $\cos\theta$ distribution of $\Upsilon(5S)\to \Upsilon(2S)\pi^+\pi^-$
consistent with the experimental data \cite{Abe:2007tk}. This observation provides an indirect evidence to the existence of two charged $Z_b$ structures, and gives a possible approach to solve the puzzles existing in the $\Upsilon(5S)$ hidden-bottom dipion decays. However, we must find the source to generate the $Z_b(10610)$ and $Z_b(10650)$ structures. In the following, we introduce the exotic state explanations to $Z_b(10610)$ and $Z_b(10650)$.

\section{Exotic state assignments to two charged $Z_b$
structures} \label{sec:2}

Since the charged $Z_b(10610)$ and $Z_b(10650)$ are close to the $B\bar{B}^*$ and $B^*\bar{B}^*$ thresholds, respectively, $Z_b(10610)$ and $Z_b(10650)$ can be as good candidate of exotic state. Before the observation of these two charged bottomonium-like states, the analysis in Refs. \cite{Liu:2008fh,Liu:2008tn} indicates that there probably exists a loosely bound S-wave $B\bar{B}^*$ molecular state, where the One-Boson-Exchange (OBE) model is applied to the dynamical calculation.

For further understanding the structure of $Z_b(10610)$ and $Z_b(10650)$, different theoretical groups have performed the study of $Z_b(10610)$ and $Z_b(10650)$ considering different exotic state assignments to $Z_b(10610)$ and $Z_b(10650)$, which mainly include molecular states composed of $B^{(*)}\bar B^{(*)}$ mesons \cite{Bondar:2011ev,Zhang:2011jja,Voloshin:2011qa,Sun:2011uh,Cleven:2011gp,Cui:2011fj}, and hidden-bottom tetraquark states \cite{Yang:2011rp,Guo:2011gu,Navarra:2011xa}.

By the OBE model, we systematically calculated the interaction between $B^{(*)}$ and $\bar{B}^{(*)}$ mesons, where $\pi$, $\rho$, $\omega$, $\sigma$ meson exchanges are introduced when deducing the effective potential of $B^{(*)}\bar{B}^{(*)}$ molecular systems. If $Z_b(10610)$ and $Z_b(10650)$ are $B\bar{B}^*$ and $B^*\bar{B}^*$ molecular states respectively, we can construct their flavor wave functions  \cite{Liu:2008fh,Liu:2008tn},
$|{Z_b(10610)}^\pm\rangle=\frac{1}{\sqrt{2}}\big(|B^{*\pm}\bar{B}^0\rangle+|B^\pm\bar{B}^{*0}\rangle\big)$ and
%|{Z_b(10610)}^-\rangle=\frac{1}{\sqrt{2}}\big(|B^{*-}\bar{B}^0\rangle+|B^-\bar{B}^{*0}\rangle\big),\\
$|{Z_b(10610)}^0\rangle=\frac{1}{2}\Big[\big(|B^{*+}B^-\rangle-|B^{*0}\bar{B}^0\rangle\big)
+\big(|B^+B^{*-}\rangle-|B^0\bar{B}^{*0}\rangle\big)\Big]$ for $Z_b(10610)$,
%|{Z_b(10650)}^-\rangle=|B^{*-}\bar{B}^{*0}\rangle\\
$|{Z_b(10650)}^0\rangle=\frac{1}{\sqrt{2}}\big(|B^{*+}B^{*-}\rangle-|B^{*0}\bar{B}^{*0}\rangle\big)$ and $|{Z_b(10650)}^\pm\rangle=|B^{*\pm}\bar{B}^{*0}\rangle$ for $Z_b(10650)$.

Under the Breit approximation, the effective potential of
the $B\bar{B}^*$ and $B^*\bar{B}^*$ systems can be related to the scattering amplitude
\begin{eqnarray*}
\mathcal{V}_E^{B^{(*)}\bar{B}^{(*)}}(\mathbf{q})&=&-\frac{\mathcal{M}({B^{(*)}\bar{B}^{(*)}}\to
{B^{(*)}\bar{B}^{(*)}})}{\sqrt{\prod_i 2M_i \prod_f 2M_f}},
\end{eqnarray*}
where $M_{i}$ and $M_j$ denote the masses of the initial and final
states respectively. Thus, we obtain the potential in the coordinate space
$\mathcal{V}(\mathbf{r})$ by performing the Fourier
transformation
\begin{eqnarray}
\mathcal{V}_E^{B^{(*)}\bar{B}^{(*)}}(\mathbf{r})=\int\frac{d\mathbf{p}}{(2\pi)^3}\,e^{i
\mathbf{p}\cdot
\mathbf{r}}\mathcal{V}_E^{B^{(*)}\bar{B}^{(*)}}(\mathbf{q})\mathcal{F}^2(q^2,m_E^2),
\end{eqnarray}
where the monopole form factor (FF)
$\mathcal{F}(q^2,m_E^2)=({\Lambda^2-m_E^2})/({\Lambda^2-q^2})$ is introduced, which
reflects the structure effect of the vertex of the heavy mesons
interacting with the light mesons. $m_E$ denotes the exchange
meson mass. We consider both S-wave and D-wave interactions
between $B^{(*)}$ and $\bar{B}^{(*)}$ mesons, which make the
$B\bar{B}^{*}$ and $B^*\bar{B}^{*}$ to be further expressed as
\begin{eqnarray}
\big|{Z_{B\bar{B}^*}^{(\alpha)}}^{(\prime)}\big\rangle&=&\left
(\begin{array}{c}
\Big|BB^*(^3S_1)\Big\rangle\\
\Big|BB^*(^3D_1)\Big\rangle\end{array}\right ),\,
\big|Z_{B^*\bar{B}^*}^{(\alpha)}[\mathrm{0}]\big\rangle=\left
(\begin{array}{c}
\Big|B^*\bar{B}^*(^1S_0)\Big\rangle\\
\Big|B^*\bar{B}^*(^5D_0)\Big\rangle,
\end{array}\right )\nonumber\\
\big|Z_{B^*\bar{B}^*}^{(\alpha)}[\mathrm{1}]\big\rangle&=&\left
(\begin{array}{c}
\Big|B^*\bar{B}^*(^3S_1)\Big\rangle\\
\Big|B^*\bar{B}^*(^3D_1)\Big\rangle\\
\Big|B^*\bar{B}^*(^5D_1)\Big\rangle\end{array}\right ),\,
\big|Z_{B^*\bar{B}^*}^{(\alpha)}[\mathrm{2}]\big\rangle=\left
(\begin{array}{c}
\Big|B^*\bar{B}^*(^5S_2)\Big\rangle\\
\Big|B^*\bar{B}^*(^1D_2)\Big\rangle\\
\Big|B^*\bar{B}^*(^3D_2)\Big\rangle\\
\Big|B^*\bar{B}^*(^5D_2)\Big\rangle\end{array}\right )\nonumber\\
\label{11}
\end{eqnarray}
with $\alpha=S,T$, where we use the notation $^{2S+1}L_J$ to
denote the total spin $S$, angular momentum $L$, total angular
momentum $J$ of the $B\bar{B}^*$ or $B^*\bar{B}^*$ system. Indices
$S$ and $D$ indicate that the couplings between $B^*$ and
$\bar{B}^*$ occur via the $S$-wave and $D$-wave interactions,
respectively.

In general, the total effective potentials of the $B\bar{B}^*$ and
$B^*\bar{B}^*$ systems are
\begin{eqnarray}
V_{\mathrm{Total}}^{{Z_{B\bar{B}^*}^{(\alpha)}}^{(\prime)}}&=&\Big\langle {Z_{B\bar{B}^*}^{(\alpha)}}^{(\prime)}\big)\Big| \sum_{E=\mathrm{\pi,\eta,\sigma,\rho,\omega}}\mathcal{V}_E^{B\bar{B}^*}(r)\big|{Z_{B\bar{B}^*}^{(\alpha)}}^{(\prime)}\big\rangle,\\
V_{\mathrm{Total}}^{{Z_{B^*\bar{B}^*}^{(\alpha)}}[\mathrm{J}]}&=&\Big\langle
{Z_{B^*\bar{B}^*}^{(\alpha)}}[\mathrm{J}]\big)\Big|
\sum_{E=\mathrm{\pi,\eta,\sigma,\rho,\omega}}\mathcal{V}_E^{B^*\bar{B}^*}(r)\big|{Z_{B^*\bar{B}^*}^{(\alpha)}}[\mathrm{J}]\big\rangle,
\end{eqnarray}
which are $2\times 2$ and $(J+2)\times (J+2)$ matrices,
respectively (see Ref. \cite{Sun:2011uh} for more details).

\begin{table}[htb]
\caption{The obtained bound state solutions (binding energy $E$
and root-mean-square radius $r_{\mathrm{RMS}}$) for the
$Z_b(10610)$ and $Z_b(10650)$ systems \cite{Sun:2011uh}.\label{BBS-1}}
\centering
\label{tab:1}       % Give a unique label
\begin{tabular}{llllllll}
\hline\noalign{\smallskip}
state & $\Lambda$ (GeV) & $E$ (MeV) & $r_{RMS}$ (fm)&state & $\Lambda$ (GeV) & $E$ (MeV) & $r_{RMS}$ (fm)  \\[3pt]
\tableheadseprule\noalign{\smallskip}
&2.1&-0.22&3.05& &2.2&-0.81&1.38\\
$Z_{b}(10610)$&2.3&-1.64&1.31& $Z_b(10650)$&2.4&-3.31&0.95\\
&2.5&-4.74&0.84& &2.8&-14.94&0.52
\\
\noalign{\smallskip}\hline
\end{tabular}
\end{table}

The numerical result listed in Table \ref{tab:1} indicates that $Z_b(10610)^\pm$ and $Z_b(10650)^\pm$ can be explained as the $B^*\bar{B}$ and $B^*\bar{B}^{*}$ molecular states \cite{Sun:2011uh}.

\section{Initial single pion emission mechanism}\label{sec:3}

Besides explaining $Z_b(10610)$ and $Z_b(10650)$ under the exotic state assignments, we also explored whether there exist other mechanisms resulting in $Z_b(10610)$ and $Z_b(10650)$ enhancement structures, which is an interesting research topic. Since the observation of the $Z_b(10610)$ and $Z_b(10650)$ structures is from the hidden-bottom dipion decays of $\Upsilon(5S)$, studying the decay mechanisms existing in the $\Upsilon(5S)$  decays can provide some hints to understand $Z_b(10610)$ and $Z_b(10650)$.

\begin{figure*}[htb]
\centering
\includegraphics[width=0.23\textwidth]{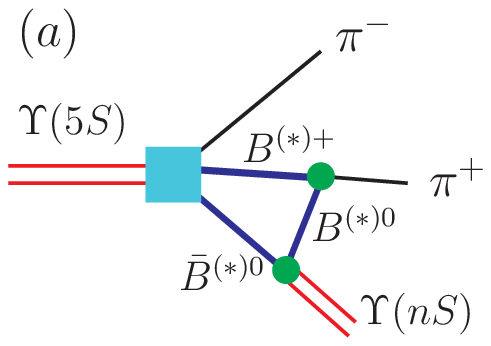}\qquad\qquad\includegraphics[width=0.23\textwidth]{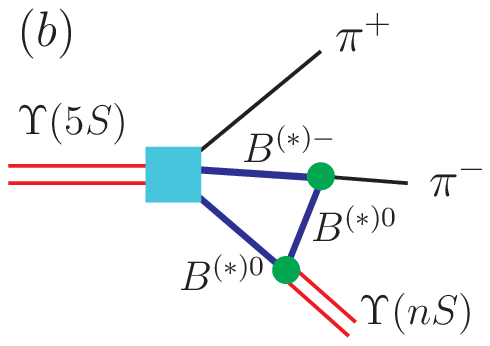}
\caption{(Color online.) The schematic diagrams describing the ISPE mechanism of $\Upsilon(5S)$. Here, we use $\Upsilon(5S)\to \Upsilon(nS)\pi^+\pi^-$ as an example. }
\label{fig:3}       % Give a unique label
\end{figure*}

Just presented in Sect. \ref{sec:2}, there are at least three different decay mechanisms just shown in Fig. \ref{fig:1}. Apart from the first two mechanisms corresponding to the first two diagrams of Fig. \ref{fig:1},
we noticed that the last two diagrams in Fig. \ref{fig:1} can be naturally replacements as the diagrams in Fig. \ref{fig:3}, where the emitted single pion directly from $\Upsilon(5S)$ makes the intermediate $B^{(*)}\bar{B}^{(*)}$ meson pair with low momenta, which interact with each other to transit into final states by exchanging $B^{(*)}$ mesons. This decay mechanism is named as the Initial Single Pion Emission (ISPE) mechanism in Ref. \cite{Chen:2011pv}.

\begin{figure*}[htb]
\centering
\includegraphics[width=0.8\textwidth]{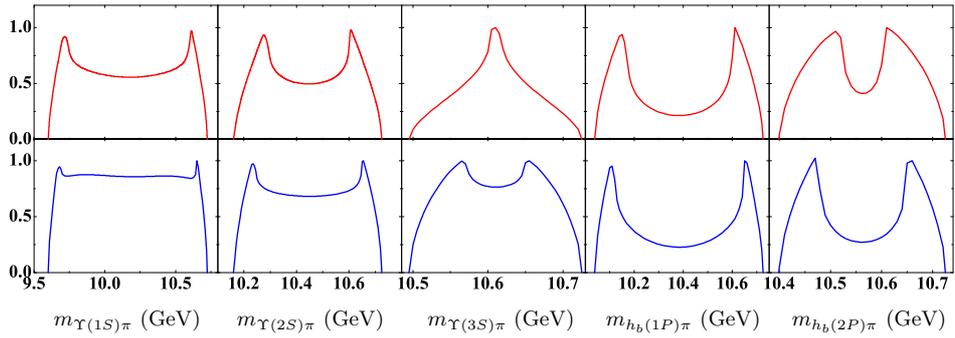}
\caption{(Color online.) The invariant mass spectra of $\Upsilon(nS)\pi^\pm$ ($n=1,2,3$) and $h_b(mP)\pi^\pm$ ($m=1,2$) of $\Upsilon(5S)\to \Upsilon(nS)\pi^+\pi^-$ and $\Upsilon(5S)\to h_b(mP)\pi^+\pi^-$ decays. }
\label{fig:4}       % Give a unique label
\end{figure*}

By the ISPE mechanism \cite{Chen:2011pv}, we studied the line shapes of ${d\Gamma(\Upsilon(5S\to
\Upsilon(nS)\pi^+\pi^-))}\over{dm_{\Upsilon(nS)\pi^+}}$ ($n=1,2,3$) and
${d\Gamma(\Upsilon(5S\to h_b(mP)\pi^+\pi^-))}\over{dm_{h_b(mP)\pi^+}}$
($m=1,2$). We found sharp structures around 10610 MeV and 10650 MeV
in the obtained theoretical line shapes of ${d\Gamma(\Upsilon(5S\to
\Upsilon(nS)\pi^+\pi^-))}\over{dm_{\Upsilon(nS)\pi^+}}$ and
${d\Gamma(\Upsilon(5S\to h_b(mP)\pi^+\pi^-))}\over{dm_{h_b(mP)\pi^+}}$
distributions, which could naturally correspond to the
$Z_b(10610)$ and $Z_b(10650)$ structures newly observed by Belle \cite{Collaboration:2011gj}. In Fig. \ref{fig:4}, we shown the numerical result in the scenario of ISPE.

If the ISPE mechanism is a universal mechanism existing the decay of heavy flavor quarkonia, we extend this mechanism to study the hidden-charm dipion decays of higher charmonia \cite{Chen:2011xk}. We predicted the charged charmonium-like structures around the $D\bar{D}^*$ and $D^*\bar{D}^*$ thresholds. Since these novel phenomena are accessible by the BESIII, Belle, BaBar experiments, and Belle-II or SuperB, we suggest future experiments to carry out the search for these charged charmonium-like structures.
\begin{figure*}[htb]
\centering
$Y(4040)$\qquad\qquad\qquad\qquad\qquad\qquad\qquad\qquad$Y(4160)$\\
\includegraphics[width=0.45\textwidth]{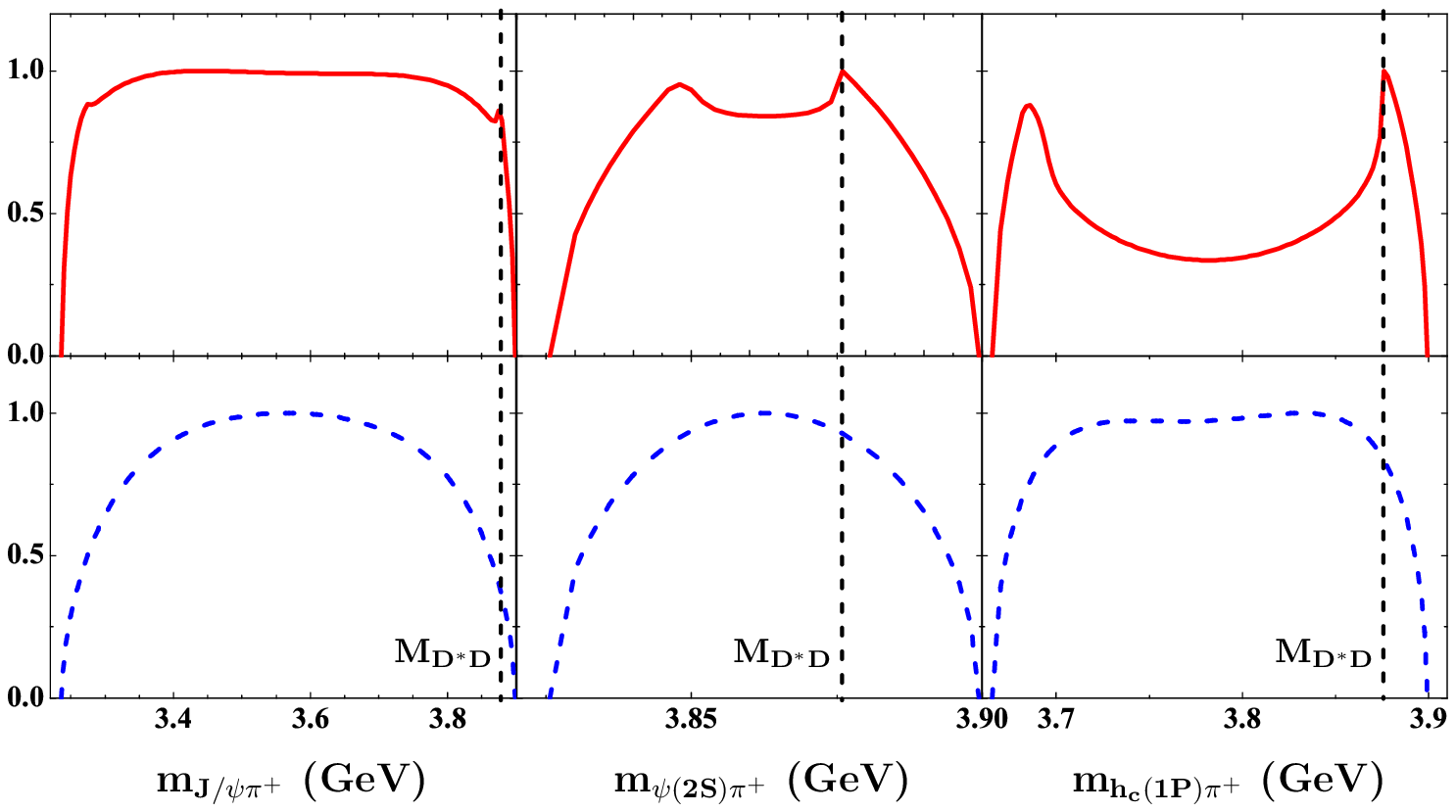}\includegraphics[width=0.45\textwidth]{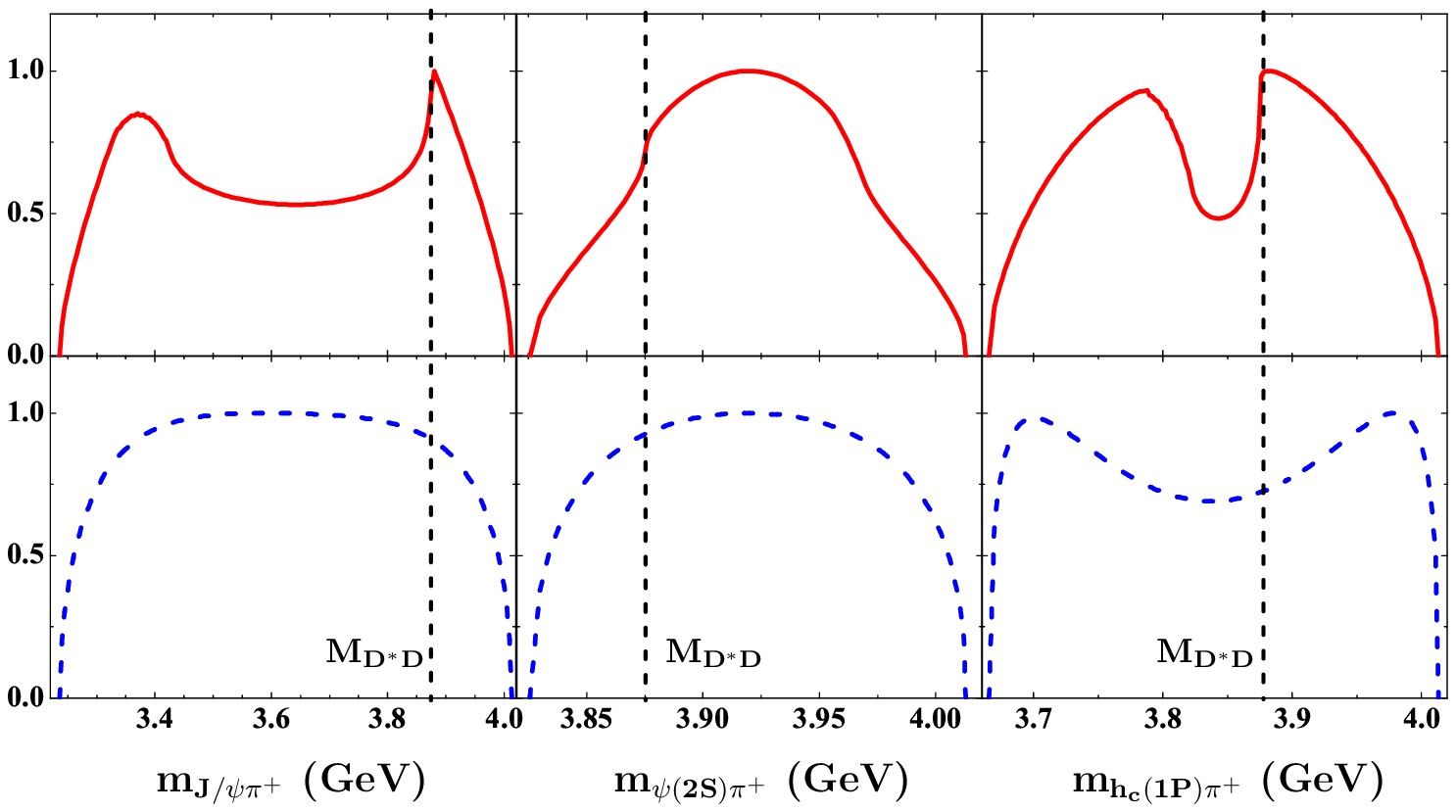}\\
$Y(4260)$\qquad\qquad\qquad\qquad\qquad\qquad\qquad\qquad$Y(4415)$\\
\includegraphics[width=0.45\textwidth]{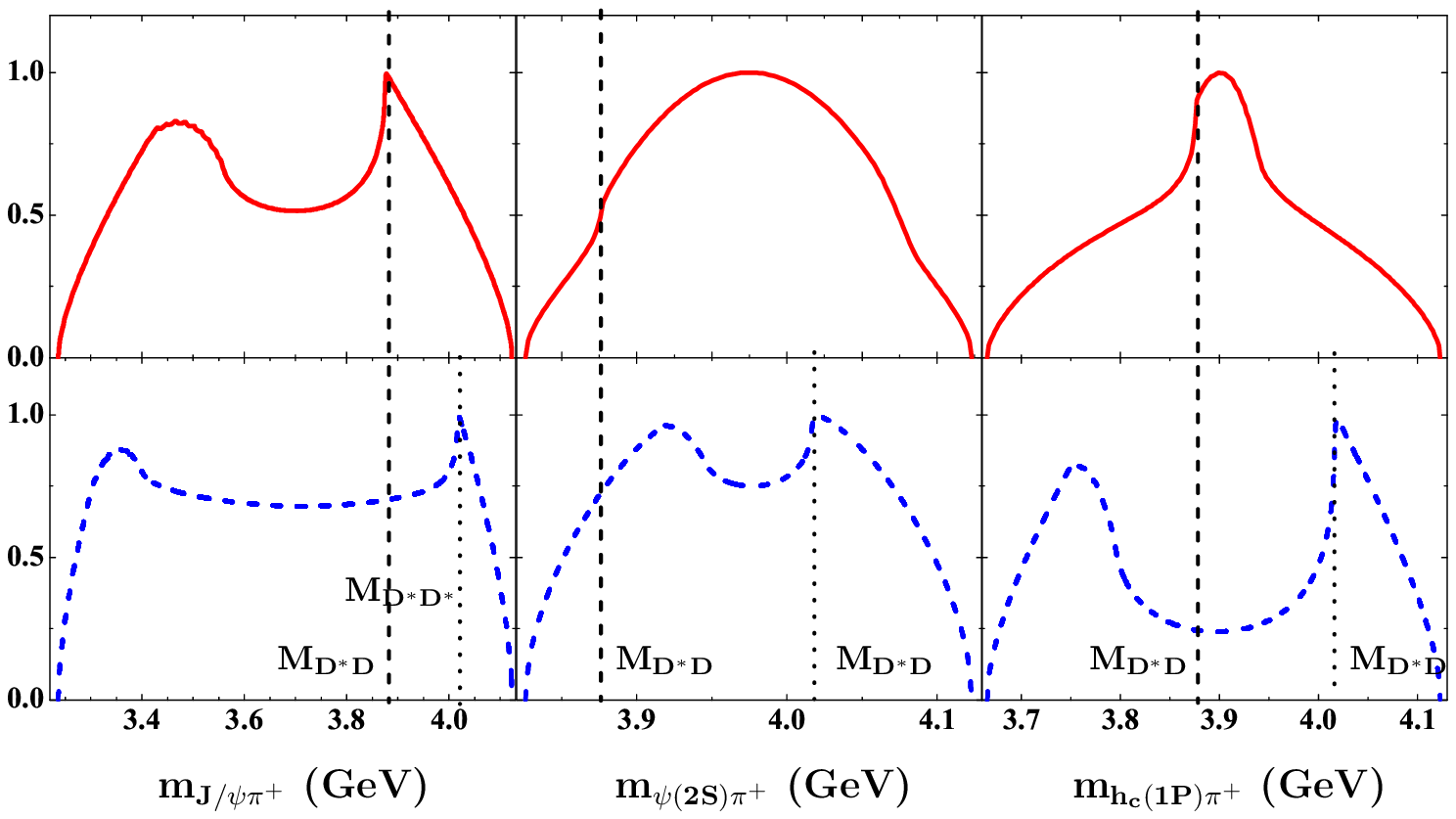}\includegraphics[width=0.45\textwidth]{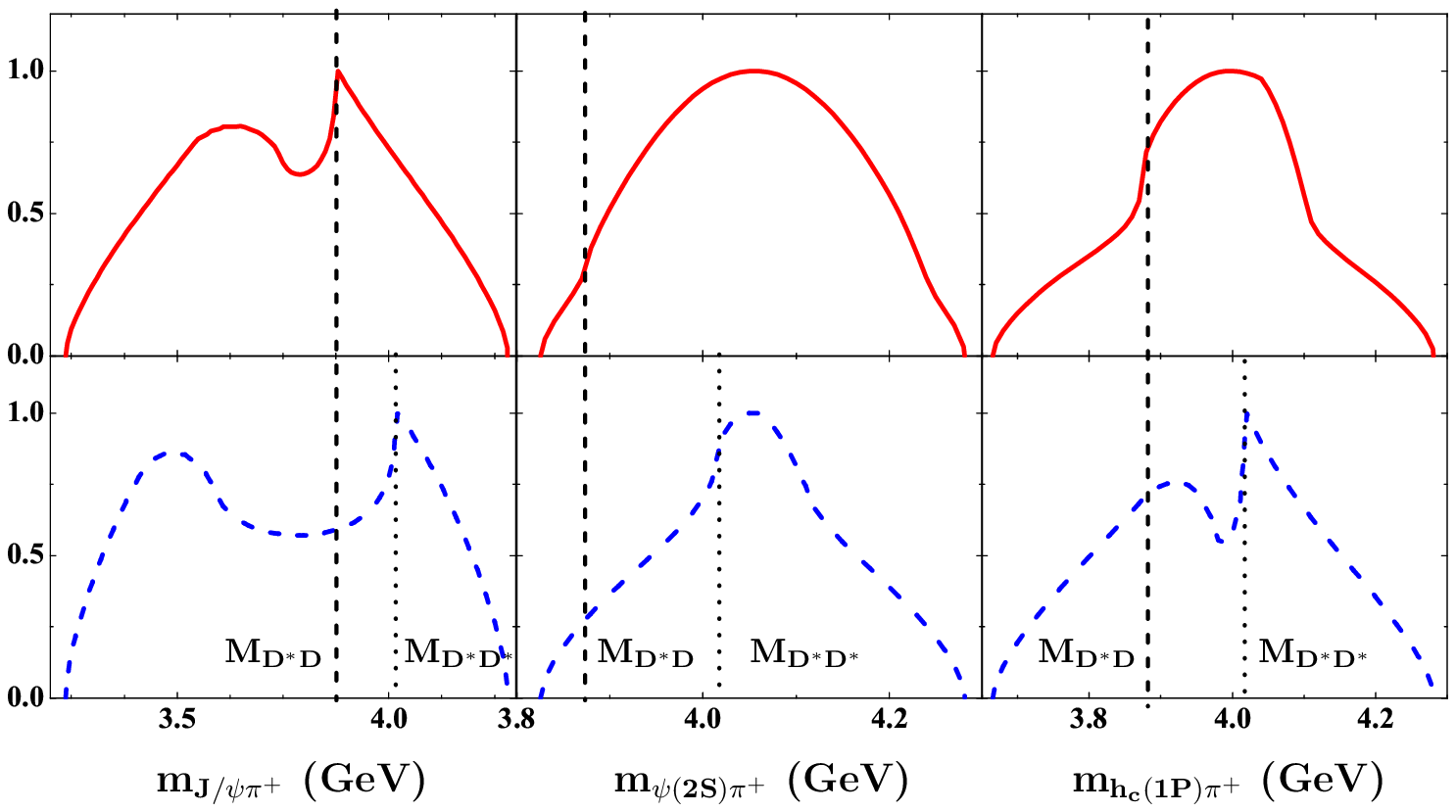}
\caption{(Color online.) The predicted invariant mass spectra of
$J/\psi\pi^+$, $\psi(2S)\pi^+$ and $h_c(1P)\pi^+$ for the
$\psi(4040)$, $\psi(4160)$, $\psi(4415)$ and $Y(4260)$ decays into
$J/\psi\pi^+\pi^-$, $\psi(2S)\pi^+\pi^-$ and $h_c(1P)\pi^+\pi^-$.
Here, the solid, dashed correspond to the results considering
intermediate  $D\bar{D}^*+h.c.$ and $D^*\bar{D}^*$ respectively in
Fig. \ref{fig:3}. The vertical dashed lines and the dotted lines
denote the threshhold of $D^\ast \bar{D}$ and $D^\ast
\bar{D}^\ast$ respectively \cite{Chen:2011pv}. Here, the maximum of the line shape is normalized to 1. }
\label{fig:5}       % Give a unique label
\end{figure*}

\section{Summary}\label{sec:4}

Recently two charged bottomonium-like structures $Z_b(10610)$ and $Z_b(10650)$ were announced by the Belle Collaboration \cite{Collaboration:2011gj}, which inspired theorists' interest in understanding their structure and underlying mechanism behind these novel phenomena. At present, different explanations to $Z_b(10610)$ and $Z_b(10650)$ were proposed due to the peculiarities of $Z_b(10610)$ and $Z_b(10650)$.

We briefly introduce the research progress on the study of $Z_b(10610)$ and $Z_b(10650)$. Especially, we
emphasize our theoretical work. In Ref. \cite{Chen:2011zv}, we indicated that two $Z_b$ structures are important to understand the anomalous $\cos\theta$ distribution of $\Upsilon(5S)\to \Upsilon(2S)\pi^+\pi^-$.
$B\bar{B}^*$ and $B^*\bar{B}^*$ molecular state assignments provide the possible explanation to the structure of $Z_b(10610)$ and $Z_b(10650)$. In addition, we found the ISPE mechanism existing $\Upsilon(5S)$ hidden-bottom dipion decays \cite{Chen:2011xk}, which results in the enhancement structures around the $B\bar{B}^*$ and $B^*\bar{B}^*$ thresholds. This observation also provides a possibility to understand $Z_b(10610)$ and $Z_b(10650)$. What is more important it that we also predicted charged charmonium-like structures close to
the $D\bar{D}^*$ and $D^*\bar{D}^*$ thresholds. The further experimental search for these structures will be helpful to test the ISPE mechanism proposed in Ref. \cite{Chen:2011xk}.

In conclusion, we still need to pay more theoretical and experimental efforts to reveal the underlying
mechanism relevant to the charged bottomonium-like structures $Z_b(10610)$ and $Z_b(10650)$.

\section{Acknowledgement}

X.L. would like to thank the organizers of APFB2011 conference,
especially Prof. Hyun-Chul Kim for his warm invitation. X.L. also
thank Lanzhou University for financial support (from the 985
Project). We enjoy the collaboration with Prof. Shi-Lin Zhu, Prof. Xue-Qian Li, Dr. Jun He, Zhi-Gang Luo and Zhi-Feng Sun. This project is partly supported by the National Natural
Science Foundation of China under Grants No. 11175073, No 11005129, No. 11035006, No. 11047606; the Ministry of Education of
China (FANEDD under Grant No. 200924, DPFIHE under Grant No.
20090211120029, NCET under Grant No. NCET-10-0442, the Fundamental
Research Funds for the Central Universities); and the West
Doctoral Project of Chinese Academy of Sciences.

%\begin{acknowledgements}
%\end{acknowledgements}

% BibTeX users please use
%\bibliographystyle{spbasic}


\end{document}